# Liquid-like thermal conduction in a crystalline solid

B. Li[1,*], Y. Kawakita[1], Q. Zhang[2], H. Wang[3,*], M. Feygenson[4], H. L. Yu[5], D. Wu[6], K. Ohara[7], T. Kikuchi[1], K. Shibata[1], T. Yamada[8], Y. Chen[5], J. Q. He[6,*], D. Vaknin[2], R. Q. Wu[3], K. Nakajima[1] and M. G. Kanatzidis[9,*]

[1]*J-PARC Center, Japan Atomic Energy Agency, Tokai, Ibaraki 319-1195, Japan.*

[2]*Ames Laboratory and Department of Physics and Astronomy, Iowa State University, Ames, Iowa 50011, USA.*

[3]*Department of Physics and Astronomy, University of California, Irvine, California 92697, USA.*

[4]*Jülich Center for Neutron Science, Forschungszentrum Jülich GmbH, D-52425, Jülich, Germany.*

[5]*Department of Mechanical Engineering, The University of Hong Kong, Pokfulam Road, Hong Kong SAR, China.*

[6]*Department of Physics, South University of Science and Technology of China, Shenzhen 518055, China.*

[7]*SPring-8, Japan Synchrotron Radiation Research Institute, Sayo, Hyogo 679-5198, Japan.*

[8]*Neutron Science and Technology Center, Comprehensive Research Organization for Science and Society (CROSS), Tokai, Ibaraki 319-1106, Japan.*

[9]*Department of Chemistry, Northwestern University, Evanston, Illinois 60208, USA.*

*Corresponding authors. Email: bing.li@j-parc.jp (B. L.); huiw2@uci.edu (H. W.); he.jq@sustc.edu.cn (J. Q. H.); m-kanatzidis@northwestern.edu (M. G. K.).

**As a generic property, all substances transfer heat through microscopic collisions of constituent particles[1]. A solid conducts heat through both transverse and longitudinal acoustic phonons, but a liquid employs only longitudinal vibrations[2,3]. As a result of this difference, a solid is usually thermally more conductive than a liquid. In canonical viewpoints, such a difference also acts as the dynamic signature distinguishing a solid from a liquid. Here, we report liquid-like thermal conduction observed in the crystalline solid $AgCrSe_2$. The transverse acoustic (TA) phonons that are exclusively dominated by motions of Ag atoms are completely suppressed by the ultrafast dynamic disorder while only the**

**longitudinal acoustic (LA) phonon takes effect in thermal transport. This microscopic insight might reshape the fundamental understanding on thermal transport properties of matter and opens up an emergent opportunity to optimize performances of thermoelectrics.**

Thermal transport is one of the most fundamental properties of matter[1]. As a persistent challenge, rationalizing anomalous thermal transport properties has been reforming our understanding on solid materials, such as the birth of the revolutionizing concept of phonon glasses[4-6] and the wave-particle duality of phonon transport in superlattices[7]. Meanwhile, materials with small thermal conductivity are desirable for a wide variety of applications like thermal insulation[8], optical phase-change memory devices[9], and efficient thermoelectric energy conversion[10]. Therefore, it is of both fundamental and technological interests to explore suppressed thermal transport in solids. By conducting a comprehensive study combining state-of-the-art neutron/X-ray scattering with density-functional-perturbation-theory (DFPT) calculations on the chalcogenide $AgCrSe_2$, we found the low-lying intense TA phonons that are dominated by motions of Ag atoms competes with fluctuations inherent in the order-to-disorder transition of Ag occupation. Phonons are significantly damped as temperature rises. The TA phonons completely collapse above the transition temperature where their *lifetime* is shorter than the relaxation time of local fluctuations while the LA mode remains, appearing liquid-like thermal conduction with extremely low lattice thermal conductivity ($\kappa_L$), ~ 0.2 $Wm^{-1}K^{-1}$ at 500 K[11-13].

$AgCrSe_2$ crystallizes in a hexagonal structure with alternative Ag layers and $CrSe_6$ octahedral layers repeating along the *c* axis[14], as shown in **Fig. 1a**. Ag atoms lie in the equivalent tetrahedral interstitial sites (I and II) in the *van der Waals gap* between $CrSe_6$ layers. The total energy calculations confirm this degeneracy (Table S3). These two sites form a buckled honeycomb lattice perpendicular to the *c* axis. In the ground state, it is expected that only one specific site (here, let's distinguish as I) is fully occupied. As temperature rises, an increasing number of Ag ions immigrate to sites II owing to the jump diffusion. At $T_c$ of about 450 K, the occupation of Ag atoms undergoes an order-to-disorder transition to the high-temperature phase with 50% occupation in each site while the space group of crystal symmetry changes from $R3m$ to $R\bar{3}m$. This transition is also regarded as a superionic transition due to the noticeable increase of ionic conduction[15]. The crystallographic changes are evidenced at X-ray scattering structure factor $S^X(Q)$ by the disappearance of Bragg peaks (003) and (006) as well as the weakening of $(10\bar{5})$, as shown in the inset of **Fig. 1c**. Neutron powder diffraction was used in tandem to track the evolution. The power-law fitting of the integrated intensity of the $(10\bar{5})$ Bragg peak gives rise to the critical exponent $\beta$ of 0.16(2) (Fig. S1), quite close to the theoretical one for the two-dimensional Ising model (0.125)[16].

At finite temperatures, the occupational disorder results in a strong diffuse scattering centered at about 2.0 Å$^{-1}$ (~ $Q_{100}$) (**Fig. 1c**). The diffuse scattering is understood in real space by Fourier-transforming $S^X(Q)$ into pair distribution function (PDF), $G^X(r)$. Shown in **Fig. 2a** are $G^X(r)$ at selected temperatures. In the ordered crystal model (Fig. S5), the first peak located at about 2.5 Å is the superposition of the nearest neighboring Cr-Se (octahedral coordination), Ag-Se (tetrahedral coordination), and Ag-Cr correlations. The second peak involves nearest neighboring Cr-Cr, Se-Se and Ag-Ag correlations, whose separations distribute around the lattice constant $a$ (3.66339 Å at 341 K). In reality, however, the uniform nearest neighboring Ag-Ag distance is split into three sets due to the occupational disorder (see **Fig. 1b**) and the next nearest neighboring Ag-related correlations sequentially become diverse as well. This is responsible for the special temperature dependence of $G^X(r)$. With the guide of the partial PDF of Ag-related correlations shown in the underpart of **Fig. 2a**, it can be seen that heights of Ag-related peaks are much more susceptible to the change of temperatures, such as those at 4.5, 13 and 19 Å (highlighted by vertical shaded bars). As a representative, the integrated intensity of the peak at 4.5 Å is plotted in **Fig. 2b**, which exhibits a well-defined critical-like behavior, in contrast to the 3.5 Å peak where contributions of Ag-related pairs are marginal. Since the integrated intensity of a PDF peak is linked to the coordination number of associated pairs[17], the decreased intensity suggests that Ag-related pairs gradually lose their coordination with approaching the transition.

The experimental $G^X(r)$ is fitted to the $R3m$ crystal model. Shown in **Fig. 2c** is the comparison at 623 K. The discrepancies are found to be associated with Ag-related pairs. The model produces higher intensity because the actual disorder effect decreases the coordination. In this sense, the difference between this model and experimental data actually defines the degree of disorder, which is quantitatively evaluated by the goodness $R_w$ generated in the fitting. The temperature dependence of $R_w$ is obtained by applying similar fitting to data collected at other temperatures, as plotted in **Fig. 2d**. It stays around 0.14 at lower temperatures, whereas tends to saturate after a fast growth in the vicinity of $T_c$. This is in agreement with the critical-like behavior of the integrated intensity shown in **Fig. 2b**. They together suggest the complete occupational disorder of Ag above $T_c$.

The thermally populated occupational disorder against the uniform long-range structure manifests itself dynamically as the disorder-phonon coupling. Inelastic neutron scattering is an ideal approach in this aspect, with which scattering cross-section of phonons can be probed throughout the Brillouin zones and meanwhile the momentum dependent energy scales of diffuse scattering can also be mapped out. The dynamic structure factor $S(Q,E)$ obtained at 5 K, as a function of momentum transfer $Q$ and energy transfer $E$, is plotted in **Fig. 3a**. Far away from the intense elastic line, there exist four less intense bands centered at about 3.5, 10, 13, and 18 meV (**Fig. 3c**). Note that the intensity spreading from $Q \sim 2.2$ Å$^{-1}$ and persisting up to 20 meV is attributed to magnons of $Cr^{3+}$ spins. To rationally assign these modes, the phonon dispersions (**Fig. 3b**) and phonon density of state (PDOS) (**Fig. 3e**) are calculated by using the DFPT approach. We identify that peaks at about 3.5, 10, 13, and 18 meV correspond to the energies of TA, LA, longitudinal optical (LO) and transverse optical (TO) modes near Brillouin-zone boundaries, respectively. Dispersive-like intensity emanating from the Bragg peak (00$\underline{15}$) and ceasing at about 3.5 meV verifies that the mode of 3.5 meV is of acoustic phonons in nature (see Fig. S6). Due to the lack of complete lattice dynamics calculations and high-resolution data, Damay et al. attributed this mode to a local one[12]. The TA phonons are dominated by motions of Ag (the integrated area of PDOS is about 90% of the total) while the TO mode mostly arises from motions of Se. This peculiar PDOS might be attributed to the distinct masses of the constituent atoms and weaker bonding between Ag and Se across the *van der Waals gap*. Indeed, our calculation on isostructural $LiCrSe_2$ indicates motions of Se are predominantly involved in its TA modes whereas the Li-related modes are pushed up to 50 meV (Fig. S12).

As the phonon softening is widely observed in many superionics with increasing temperature[18], the energies of TA ($E_{TA}$) and TO ($E_{TO}$) modes near Brillouin-zone boundaries in $AgCrSe_2$ are remarkably decreased as shown in **Fig. 3d**. Especially, $E_{TA}$ displays a very rapid drop near $T_c$, which indicates that the softening is associated with the occupational disorder of Ag. To corroborate the relevance of the occupational disorder, phonon energies are considered in the quasi-harmonic approximation (QHA) paradigm where they are solely dependent on the interatomic distance, i.e., the volume[19]. We obtain the volume dependence of phonon dispersions (**Fig. 3b**), PDOS (**Fig. S11**), $E_{TA}$ and $E_{TO}$ (**Fig. 3f**). The phonon energies vary approximately linearly with the volumetric expansion. The phonon softening is tremendously underestimated in QHA calculations. Experimentally, for example, $E_{TA}$ is reduced by 0.5 meV when temperature increases from 150 to 400 K, where the volume expands by about 0.75% (**Fig. S3**). To explain such softening, however, an unpractically large volumetric expansion (~10%) is necessitated in QHA calculations. Furthermore, QHA calculations yield a moderate Grüneisen parameter ($\gamma$) of 0.776, indicative of weak anharmonicity in this system. In addition, the softening is not compatible with a soft-mode phase transition (**Fig. S7**). Consequently, the observed phonon softening is most likely related to the superionic behavior that results in the gradual melting of Ag sublattice and weaker interlayer coupling.

Now, we turn our attention to the dynamic aspect of the diffuse scattering. **Fig. 4a** and **b** show $S(Q,E)$ with incident energy of 5.931 meV at 150 and 520 K, respectively. It can be seen that pronounced diffuse scattering exists at $Q \sim 2.0$ Å$^{-1}$. At 150 K, the diffuse scattering appears just in the vicinity of the elastic line and TA phonons are fairly sharp. Nevertheless, the diffuse scattering becomes dominant at 520 K at the expense of Bragg peaks and TA phonons. The constant-$Q$ cutting spectra at [2.11, 2.16] Å$^{-1}$, just in between two Bragg peaks, are plotted in **Fig. 4c**. At 150 K, the profile of TA phonons is well described by a damped harmonic oscillator (DHO) function while the dynamic diffuse scattering can be fitted well by a Lorentzian function. As temperature goes up, the peak associated with TA phonons is damped (**Fig. S8**). The temperature dependencies of full width at half maximum ($\Gamma_{TA}$) is plotted in **Fig. 4d**. Above $T_c$, the spectrum is well described without the need of a DHO function. The complete suppression of TA phonons near both Brillouin-zone centers and boundaries is confirmed in the higher $Q$ region where there is fewer component of the diffuse scattering (**Fig. S6**). Similar to the strong temperature dependence of $\Gamma_{TA}$, the full width at half maximum of the diffuse scattering, $\Gamma_{Diffuse}$, is approximately 2 meV at 150 K and becomes nearly twice wider at 520 K after an obvious increase around $T_c$ (**Fig. 4d**). The robust atomic fluctuations are the origin of large ionic conductivity even below $T_c$[15]. Differing from its strong temperature dependence, $\Gamma_{Diffuse}$ is nearly $Q$ independent at a given temperature.

The ultralow $\kappa_L$ in this system is essentially ascribed to the disorder effect. As illustrated in **Fig. 1b**, the coherent vibrations of Ag atoms are gradually damped and localized by the growing occupational disorder that significantly increases the local repulsive energy (**Fig. S9**). As a result, the lifetime of TA phonons is shortened down to about 0.4 ps ($\tau_{TA} = 2\hbar/\Gamma_{TA}$, where $\hbar$ is Planck constant) at 440 K, whose product with group velocities (Table S5) yields mean free paths as short as 4.4 Å that approaches the interatomic distances. Such a short lifetime is comparable to the relaxation time ($\tau_{Diffuse} = 2\hbar/\Gamma_{Diffuse}$) of the dynamic disorder, ~ 0.39 ps. Above $T_c$, we estimate a fictitious *lifetime* for TA phonons by linearly extrapolating the temperature dependence of $\Gamma_{TA}$. For example, "$\tau_{TA}$" is about 0.3 ps at 520 K whereas $\tau_{Diffuse}$ is 0.37 ps, from which the breakdown of TA phonons is understandable. The remaining LA phonon continues to be damped (Fig. S6) and the mechanism is perhaps compatible with the scenario of a broad rattling mode[20]. The significance of the occupational disorder is also reflected in the difference between measured and calculated $\kappa_L$. The measured $\kappa_L$ in literatures are different[11-13] and it is about 0.55 Wm$^{-1}$K$^{-1}$ at room temperature in our measurement (Fig. S13). In contrast, the DFPT calculations, without disorder effects taken into account, provide $\kappa_L$ = 8.5 Wm$^{-1}$K$^{-1}$ at 300 K and $\gamma$ of 0.776. This combination nicely follows an empirical relation, $\kappa_L \sim \gamma^{-2}$ found for most thermoelectric materials in which less disorder exists[21]. In fact, this relationship overestimates $\kappa_L$ of systems with strong disorder such as AgSbTe$_2$[22].

The phonon breakdown discovered in AgCrSe$_2$ is unprecedented and sheds new light on the understanding of thermal conduction and optimization of thermoelectrics. Conventionally, $\kappa_L$ can be efficiently reduced through enhancing scattering of acoustic phonons by other phonons[23-25], an uncorrelated or concerted rattling mode[4-6,26], interfaces and other defects[27,28]. Recently, it is assumed that the liquid-like Cu atoms in superionic Cu$_2$Se might dismiss transverse modes[29]. However, the inelastic neutron scattering study suggests the robustness of acoustic phonons, probably due to the contribution from Se[30]. A plausible scenario might be a broad rattling mode as proposed for the similar compound Cu$_3$SbSe$_3$[20]. In AgCrSe$_2$, it is extraordinarily unique that the combination of the nontrivial PDOS, the order-to-disorder transition, and close time scales of TA phonons and the dynamic disorder enables an extreme disorder-phonon coupling. The ultrafast liquid-like relaxation above $T_c$ completely suppresses TA phonons and the LA phonon only survives for thermal transport, characteristic of the liquid-like thermal conduction. In considering the vast existence of correlated disorder[31], the concept of phonon breakdown is potentially a general route to reduced $\kappa_L$.

## Methods

The details are provided in the supplemental information.

**Acknowledgment** We acknowledge the award of beam time from Spallation Neutron Source, a DOE office of Science User Facility operated by the Oak Ridge National Laboratory, via proposal IPTS-13971, from SPring-8 via proposal No. 2015B1070, and from J-PARC via proposal No. 2012P0906. Ames Laboratory is operated for the U.S. Department of Energy by Iowa State University under Contract No. DE-AC02-07CH11358. Works at University of California, Irvine were supported by DOE-BES under Grant No. DE-FG02-05ER46237 and the computer simulations were partially supported by NERSC. D.W. and J.Q.H. were supported by Technology and Innovation Commission of Shenzhen Municipality under Grant No. JCYJ20150831142508365, Natural Science Foundation of Guangdong Province under Grant No. 2015A030308001, and the leading talents of Guangdong province Program under Grant No. 00201517. H.L.Y. and Y.C. acknowledge the research computing facilities offered by ITS, HKU. We thank Dr. M. Kofu for the fruitful discussion.

**Author contributions** B. L. proposed the project. Q. Z., D. V., D. W., and J. Q. H. synthesized the samples. D. W. and J. Q. H. carried out thermoelectric measurements. M. F. performed neutron powder diffraction measurements. B. L., Y. K., and K. O. collected X-ray scattering data. B. L., Y. K., T. K., K. S., T. Y., and K. N. performed inelastic neutron scattering measurements.

H. W., H. L. Y., Y. C., and R. Q. W. performed theoretical calculations. B. L. and H. W. analyzed experimental and theoretical data, respectively. B. L., H. W., J. Q. H., and M. G. K. wrote the manuscript with discussion and input from all the coauthors.

**Supplemental Information** accompanies this paper at http://www.nature.com

**Competing financial interests** The authors declare no competing financial interests.

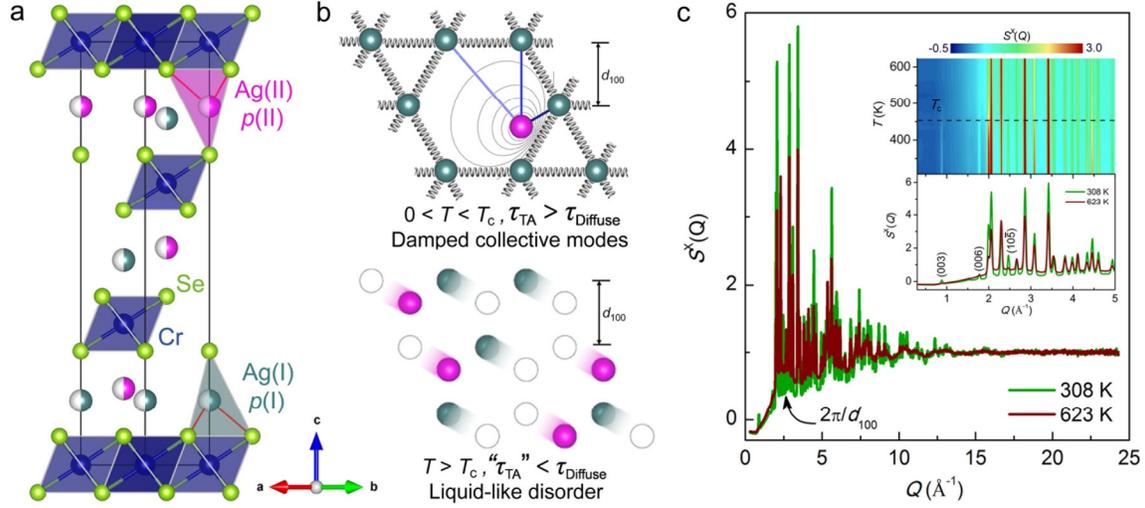

**Figure 1. Structures and phase transition. a**, The crystal structure of AgCrSe$_2$ with the CrSe$_6$ octahedra and AgSe$_4$ tetrahedra highlighted. There are two equivalent sites for Ag (I and II), whose occupations $p$(I) and $p$(II) are dependent on temperature. The phase transition is driven from $R3m$ to $R\bar{3}m$ when $p$(I) = $p$(II) at $T_c$. **b**, The schematic diagram for instantaneous structures of Ag projected along the $c$ axis. The upper panel shows that in the low-temperature phase with $p$(I) > $p$(II) the collective modes (illustrated by springs) are damped by the increase of local repulsive energy (illustrated by contour lines) induced by the occupational disorder. This energy scale is evaluated by total energy calculations shown in Fig. S9. The diverse bond lengths of three sets are labelled by solid lines. The lower panel displays the complete dynamic occupational disorder above $T_c$. The blank circles represent empty sites. Ag-lying planes are {100} and their interplanar distance $d_{100}$ is labelled in both panels. $\tau_{TA}$ is the lifetime of TA phonons while $\tau_{Diffuse}$ is the relaxation time of dynamic disorder of Ag. **c**, The structure factor, $S^X(Q)$, obtained in X-ray scattering at 308 and 623 K. The diffuse scattering appears at the scattering vector of {100} planes, about 2.0 Å$^{-1}$. The inset highlights $S^X(Q)$ at small $Q$ region for temperature evolution (upper) and two end temperatures (lower). Bragg peaks of (003), (006) and (10$\bar{5}$) are labelled. The transition temperature $T_c$ is marked at the temperature where the (003) Bragg peak tends to disappear.

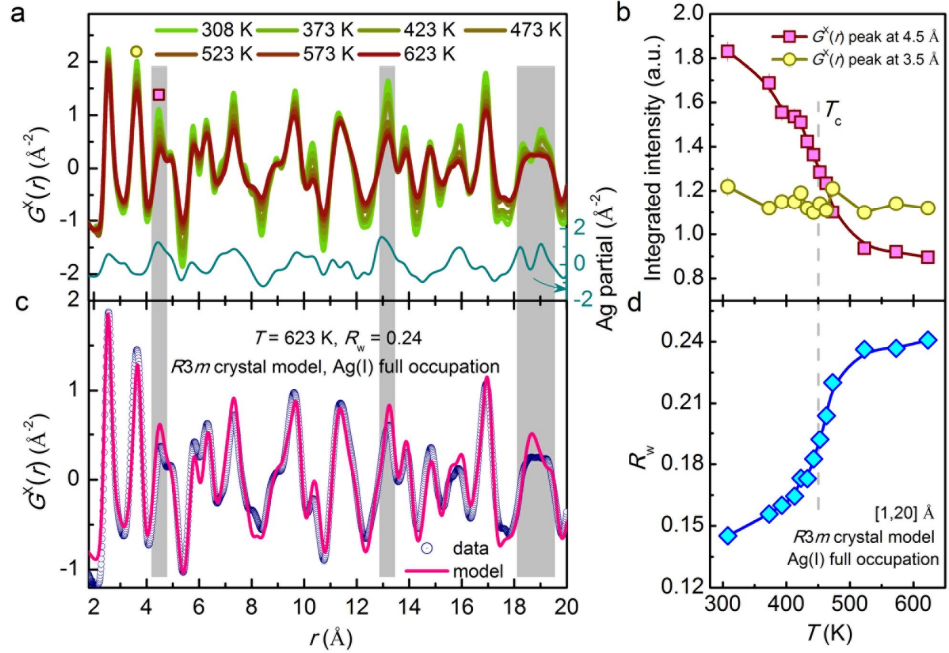

**Figure 2. Thermally populated occupational disorder of Ag atoms. a**, The experimental PDF, $G^X(r)$, obtained from X-ray scattering at selected temperatures up to 20 Å (for the comparison with neutron case, see Fig. S4). Underneath is the superposition of partial $G^X(r)$ for Ag involved pairs (Ag-Ag, Ag-Cr and Ag-Se) calculated in the $R3m$ crystal model. The partial of individual pair is shown in Fig. S5. **b**, The integrated intensity of $G^X(r)$ for Ag-correlation-poor peak at 3.5 Å and Ag-correlation-rich peak at 4.5 Å, which are labelled by a circle and a square in **a**, respectively. **c**, The real-space refinement of experimental $G^X(r)$ based on the $R3m$ crystal model at 623 K. **d**, The goodness of this real-space refinement as a function of temperature. The vertical shaded bars in **a** and **c** highlight the positions where Ag-related correlations are dominant. In **b** and **d** $T_c$ is marked by a dash line.

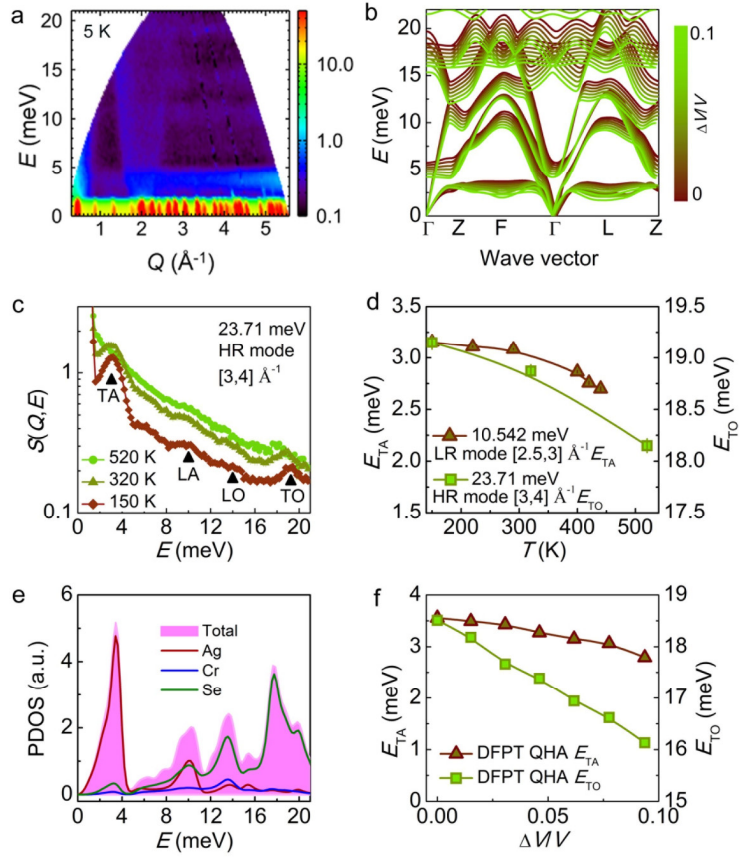

**Figure 3. Experimental and theoretical results of phonons. a**, The contour plot of dynamic structure factor $S(Q,E)$ obtained in inelastic neutron scattering with incident neutron energy $E_i$ = 23.71 meV in the low-resolution (LR) mode at 5 K. **b**, The calculated phonon dispersion relationship by using DFPT QHA method along several high symmetric directions defined in Fig. S10. The color bar depicts the magnitude of isotropic volumetric expansions that result in the monotonous softening of all modes. Those in full frequency range are shown in Fig. S11. **c**, The constant-$Q$ cuts of $S(Q,E)$ obtained at $E_i$ = 23.71 meV in the high-resolution (HR) mode in the interval of [3,4] Å$^{-1}$, where peaks correspond to energies of TA, LA, LO and TO modes near Brillouin-zone boundaries. **d**, The energies of TA and TO phonons near Brillouin-zone boundaries determined by fitting experimental spectra. **e**, Calculated PDOS. **f**, The calculated energies of TA and TO phonons defined by the peak positions at PDOS.

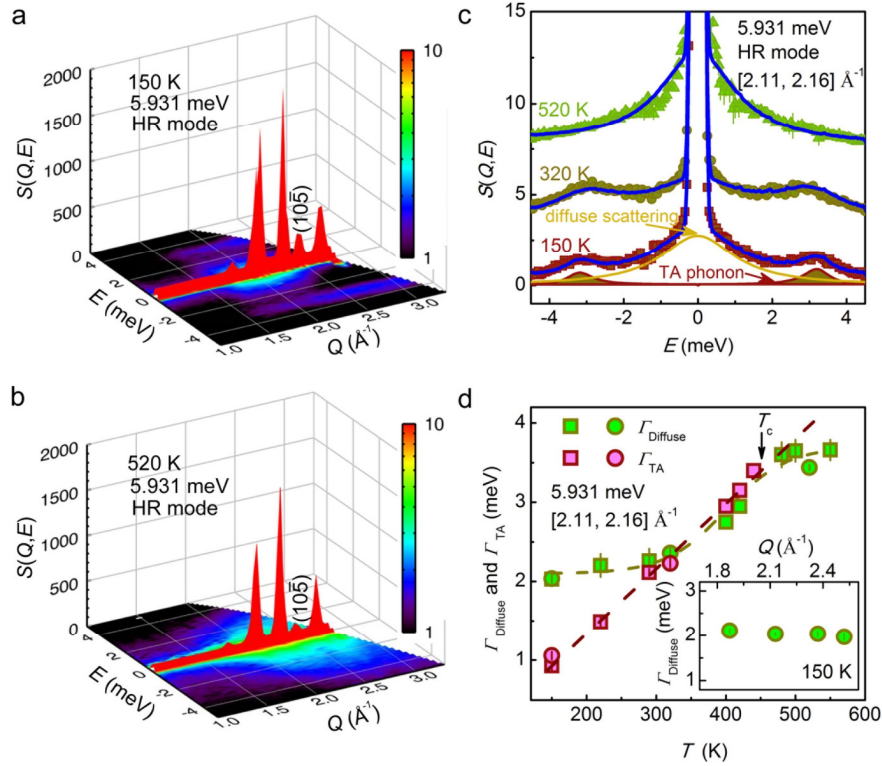

**Figure 4. Breakdown of TA phonons. a,b**, $S(Q,E)$ surface plots at 150 and 520 K at $E_i$ = 5.931 meV in the HR mode. **c**, The fitting of spectra at constant-$Q$ cuts of [2.11,2.16] Å$^{-1}$ at 150, 320, and 520 K. The symbols represent experimental data points and the solid lines are obtained from fitting. The spectra are vertically shifted for clarity. At 150 K, the DHO and Lorentzian components in the fitting are highlighted, respectively, which account for TA phonons and the dynamic diffuse scattering, respectively. **d**, The full widths at half maximum of the TA phonons and diffuse scattering. $Q$-dependence of the latter is shown in the inset. The squares and circles are data points obtained in LR and in HR modes, respectively. The dash lines are guide for eyes. $T_c$ is labelled, where $\Gamma_{TA}$ and $\Gamma_{Diffuse}$ become comparable.

# Supplemental Information for

# Liquid-like thermal conduction in a crystalline solid

B. Li*, Y. Kawakita, Q. Zhang, H. Wang*, M. Feygenson, H. L. Yu, D. Wu, K. Ohara, T. Kikuchi, K. Shibata, T. Yamada, Y. Chen, J. Q. He*, D. Vaknin, R. Q. Wu, K. Nakajima and M. G. Kanatzidis*

*Corresponding authors: bing.li@j-parc.jp (B. L.); huiw2@uci.edu (H. W.); he.jq@sustc.edu.cn (J. Q. H.); m-kanatzidis@northwestern.edu (M. G. K.).

**Including:**
**1. Methods**
**2. Supplemental figures**
**3. Supplemental tables**
**4. Supplemental references**

# Organization of the Supplemental Information

In the section of **Methods**, the details on **sample preparation**, **thermoelectric measurements**, **neutron powder diffraction**, **synchrotron X-ray scattering**, **pair distribution function analysis**, **inelastic neutron scattering**, and **theoretical calculations** are provided.

In the sections of **Supplemental figures** and **Supplemental tables**, we clarify following issues:

**a).** long-range ordered structures determined by using neutron powder diffraction: **Figure S1–S3**, and **Table S1, S2**

**b).** comparison of neutron and X-ray pair distribution functions: **Figure S4**, **S5**

**c).** softening and breakdown of the transverse acoustic phonons: **Figure S6–S8**

**d).** lattice dynamic calculations including thermal conductivity: **Figure S9–S13**, and **Table S3–S5**

In the section of **Supplemental references**, we list references involved in this supplemental information.

1. Methods

A) Sample preparation

The powder samples were prepared by using solid state reaction method[12]. The starting materials Ag, Cr and Se powder in the ratio of stoichiometric composition were ground, mixed, and pelletized. The pellets were sealed in evacuated quartz tubes. The tubes were placed in a box furnace and heated first to 473 K for 6 hours, and then to 1173 K for 24 hours, followed by the natural cooling to room temperature. Before all neutron scattering and X-ray scattering measurements, the samples were annealed under argon flow at 523 K overnight.

Spark plasma sintering (SPS) method was used to press powder into ingots for thermoelectric measurements. The applied uniaxial pressure is 60 MPa, and the sintering process was kept at 923 K for 5 minutes. The obtained ingots were cut into coins of 10 mm in radius and 2 mm in thickness, followed by fine polishing prior to thermal diffusivity measurements. The samples were coated with a thin layer of graphite to achieve good light absorption and minimize errors from variations of the samples' emissivity.

B) Thermoelectric measurements

The thermal diffusivity coefficient ($D$) was measured using the laser flash diffusivity method on a commercial Netzsch LFA457 apparatus, the thermal diffusivity data were analyzed using a Cowan model with pulse correction. The specific heat capacity ($C_p$) was determined by using a differential scanning calorimetry (Netzsch STA449, Germany). The density was determined using the Archimedes method. The thermal conductivity was then calculated from $\kappa = D \cdot C_p \cdot \rho$[S1]. The uncertainty of the thermal conductivity was estimated to be within 5%, considering the uncertainties for $D$, $C_p$ and $\rho$.

C) Neutron powder diffraction

The neutron powder diffraction measurements were performed at the Nanoscale Ordered Materials Diffractometer (NOMAD) in Spallation Neutron Source of Oak Ridge National Laboratory, USA[S2]. About 5 grams powder was sealed in a vanadium sample can under helium gas with indium wire. The constant temperature scans were taken at several temperatures by using a vacuum furnace. Each scan took about 1 hour. An empty vanadium can, a standard vanadium rod, and the background were also measured at room temperature for the pair distribution function (PDF) analysis. The integrated intensity of $(10\bar{5})$ Bragg peak was determined by fitting to the Gaussian function and the power-law fitting suggests that the transition temperature $T_c$ is determined to be 443(1) K, as shown in **Fig. S1**. The diffraction data was analyzed by using Rietveld refinement method in GSAS[S3]. The occupation of each atom was not refined. The structural model is $R3m$ blow the transition and $R\bar{3}m$ above the transition. The typical refinements are shown in **Fig. S2**. The determined lattice dimensions and the Debye-Waller factors for Ag are summarized in **Fig. S3**. The detailed crystal structure information including lattice constants, atomic coordinates, and anisotropic Debye-Waller factors are summarized in **Table S1** and **S2**.

D) Synchrotron X-ray powder scattering

The high energy X-ray powder scattering was carried out at the beam line BL04B2 of SPring-8, Japan, with photon energy of 113 keV[S4]. About 0.2 gram powder was sealed under helium gas into a quartz capillary. The constant temperature scans were taken in the vicinity of the order-to-disorder transition. Each scan took about 3.5 hours. An empty capillary sealed in helium gas was also measured at room temperature as a reference for PDF analysis.

### E) PDF analysis

$G(r)$ is obtained by Fourier transforming the normalized structure factor $S(Q)$ with cutoff, $Q_{max}$, ~ 25 Å$^{-1}$ [S5,S6]

$$G(r) = 4\pi r[\rho(r) - \rho_0] = \frac{2}{\pi}\int_0^{Qmax} Q[S(Q) - 1]\sin(Qr)dQ \quad \text{(Eq. S1)}.$$

Here, $\rho(r)$ is the microscopic pair density and $\rho_0$ is the average number density. The neutron diffraction data obtained from NOMAD was processed in PDFgetN to correct the background, multiple-step scattering, and absorption[S7]. Then, the corrected intensity was normalized to determine $S^N(Q)$. Same parameters were applied to the data at all temperatures. $S^X(Q)$ was obtained by normalizing the intensity of X-ray diffraction data collected at BL04B2 after subtraction of the background (empty capillary) and the Compton scattering. The partial $G(r)$ for each pair was calculated in PDFgui[S8] by using the crystal structure determined in neutron powder diffraction experiment at 341 K. All $G(r)$ patterns were refined by using PDFgui with the structural model of $R3m$ in the region of [1, 20] Å.

$G(r)$ is related to the bond length distribution of the material weighted by the respective scattering powers of the contributing atoms, as determined by[S6],

$$G(r) = \frac{1}{r}\sum_i\sum_j \left[\frac{b_i b_j}{\langle b\rangle^2}\delta(r - r_{ij})\right] - 4\pi r\rho_0 \quad \text{(Eq. S2)},$$

where the sum goes over all pairs of atoms $i$ and $j$ within the model crystal separated by $r_{ij}$. The scattering length of atom $i$ is $b_i$ and $\langle b\rangle$ is the average scattering length of the sample. The scattering length of Ag with respect to X-ray and neutron are very different, which leads to different weight coefficients. As compared in **Fig. S4** and **S5**, the X-ray is much better at probing Ag than neutron. It can be seen that the Ag-correlation-rich peaks have relatively higher intensity in X-ray case. The integrated intensities of $G^X(r)$ peaks at 3.5 and 4.5 Å were determined by fitting to one and two Gaussian functions, respectively.

### F) Inelastic neutron scattering

In an inelastic neuron scattering event, the incident neutron with energy $E_i$ and wave vector $\mathbf{k}_i$ is scattered by a nuclei and then change its energy and wave vector to $E_f$ and $\mathbf{k}_f$. The energy transfer $E$ and momentum transfer $\mathbf{Q}$ are defined as[S9]

$$E = \hbar\omega = E_i - E_f \quad \text{(Eq. S3)},$$

$$\mathbf{Q} = \mathbf{k}_i - \mathbf{k}_f \quad \text{(Eq. S4)}.$$

Here, $\hbar$ is Planck constant and $\omega$ is angular frequency of the motion of the nuclei. The dynamic structure factor $S(\mathbf{Q}, E)$ is the four-dimensional Fourier transform of time-dependent pair correlation function $G(\mathbf{r},t)$ of the space and time coordinates, $\mathbf{r}$ and $t$,

$$S(\mathbf{Q}, E) = \frac{1}{2\pi\hbar}\iint G(\mathbf{r},t)\exp\{i(\mathbf{Q}\cdot\mathbf{r} - \omega t)\}d\mathbf{r}dt \quad \text{(Eq. S5)}.$$

Thus, $S(\mathbf{Q},E)$ reflects the dynamic pair correlations in reciprocal space. The coherent inelastic double differential cross-section is given by

$$\frac{d^2\sigma}{d\Omega dE_f} = \langle b\rangle^2 \frac{k_f}{k_i} NS(\mathbf{Q}, E) \quad \text{(Eq. S6)}.$$

Here, $\sigma$ is scattering cross-section, $\Omega$ is solid angle, $\langle b\rangle$ is the mean scattering length, $N$ is number of nuclei. By measuring the scattering cross-section, it is allowed to understand the system in study from $S(\mathbf{Q},E)$. Multi-$E_i$ time-of-flight inelastic neutron scattering measurements were performed at the cold neutron disk chopper spectrometer BL14 AMATERAS of J-PARC in Japan[S10]. The powder sample around 7 grams was sealed in an aluminum can with indium wire. A cryostat was used to access lower temperatures and a niobium furnace was used for higher temperatures measurements. The choppers configurations were set to select $E_i$ of 23.71, 10.542, 5.931 and 2.635 meV at the low resolution (LR) mode while 23.71, 5.931 and 2.635 meV at high resolution (HR) mode[S11]. The data reduction was completed by using Utsusemi suite[S12]. The background contributed by the niobium furnace was subtracted. The resulted $S(Q,E)$ data was visualized in Mslice of DAVE[S13]. The $Q$-cutting spectra were fitted in PAN of DAVE by including a damped harmonic oscillator function[S14], a Lorentzian function, a delta function, and a constant background, which are convoluted to the instrumental resolution. They describe the transverse acoustic phonons, dynamic diffuse scattering, incoherent elastic scattering, and background, respectively. The much higher energy resolution, compared with the previous study[13], allows us to investigate spectra in great details, in particular, the interplay between the dynamic diffuse scattering and the TA phonons.

G) Theoretical calculations

a). First-principles calculations:

Density functional theory (DFT) calculations were performed with the Vienna Ab-initio Simulation Package (VASP)[S15], which implements a fully relativistic calculation for core electrons and treats valence electrons in a scalar relativistic approximation[S16,S17]. The spin-polarized generalized gradient approximation (GGA)[S18] was used for the description of the exchange-correlation interaction among electrons. We treated Ag-4d5s, Cr-3d4s and Se-4s4p as valence states and adopted the projector-augmented wave (PAW) pseudopotentials to represent the effect of their ionic cores[S19,S20]. The energy cutoff for the plane-wave expansion was 500 eV, sufficient for $AgCrSe_2$ system according to our test calculations. In this work, calculations were carried out on a 4-atom primitive rhombohedra unit cells (**Table S3**) for comparison the total energies of the two-fold degenerate states, on a $2\times2\times1$ supercell (**Fig. S9**) for estimation of energy increase due to occupational disorder, on a $1\times1\times2$ supercell (**Fig. S10** and **Table S4**) for comparison of total energies of different magnetic configurations, and on a $4\times4\times4$ supercell (**Fig. 3, S11** and **S12**) containing 256 atoms for phonons. We sampled the Brillouin zone with a $9\times9\times9$ Monkhorst-Pack[S21] k-mesh for the primitive cell and $3\times3\times3$ k-mesh for the $4\times4\times4$ supercell. Spin-orbital coupling (SOC) was not treated in the simulation due to that i) Ag, Cr, Se don't have strong SOC; ii) it greatly increases computational cost; and iii) has a minimal effect on lattice vibration[S22]. Structures were optimized with a criterion that the atomic force on each atom becomes weaker than 0.01 eV/Å and the energy convergence is better than $10^{-8}$ eV.

b). Lattice-dynamics calculations:

Density functional perturbation theory (DFPT) is a particularly powerful and flexible theoretical technique that allows calculation of electron-density linear response within the density functional framework, thereby facilitating the acquisition of vibrational frequencies within crystalline materials[S23]. Lattice dynamics calculations were carried out using the Phonopy package[S24], with VASP employed as the calculator to obtain interatomic force constants (IFCs) via a DFPT calculation, and then directly calculate the collective vibrational spectra of phonons. To balance the accuracy and computational efforts, we calculate the phonon spectra based on the ferromagnetic structure, given that energy difference between ferromagnetic and antiferromagnetic is very small (1~2 ueV/atom), as shown in **Table S4**. As for the *R*3*m* symmetry, evaluation of the IFCs requires 4×4×4 supercells which gave a good balance between accuracy and computational cost. During post-processing, sampling the phonon frequencies on a 30×30×30 centered *q* mesh converged the vibrational density of states, and hence the values of thermodynamic properties calculated from it. For the quasi-harmonic (QHA) calculations[S25], additional finite displacement calculations were performed on unit cells at approximately ±2% of the equilibrium volume in steps of 0.5%, which guarantee these volumes correspond to temperatures inside the validity range for the QHA. Having computed the volume dependence of phonon spectra of the AgCrSe$_2$ systems, we then performed further modeling on these structures, such as Debye temperature, Grüneisen parameter and thermal conductivity.

### c). Lattice thermal conductivity:

The intrinsic lattice thermal conductivity can be obtained based on Slack's model[S26], which gives

$$\kappa_L = A \frac{\bar{M}\theta_D^3 \delta}{\gamma^2 n^{2/3} T} \qquad (\text{Eq. S7})$$

where A represent a collection of physical constant ($A \approx 3.1 \times 10^{-6}$ if $\kappa_L$ is in W/mK, *M* is in amu and $\delta$ is in Å), $\bar{M}$ is the average atomic mass, $\theta_D$ is the Debye temperature, $\delta$ is the volume per atom and *n* is the number of atoms in the primitive cell. Grüneisen parameter $\gamma$ could be computed from the equation below:

$$\gamma = \frac{\xi B V_m}{C_v} \qquad (\text{Eq. S8})$$

where $\xi$ is the volumetric thermal expansion coefficient, *B* is the isothermal bulk modulus, $C_V$ the isochoric molar specific heat and $V_m$ is the molar volume. We take $\gamma$ as a temperature-independent constant and use the value at 300 K, $\gamma = 0.776$, for our lattice thermal conductivity computation that is consistent with typical thermoelectric materials[22,S26]. The large disparity between calculated and measured thermal conductivity can be ascribed to the superionic behavior. The Debye temperature is evaluated from the Debye model by fitting the low temperature heat capacity to be 177.9 K.

2. Supplemental figures

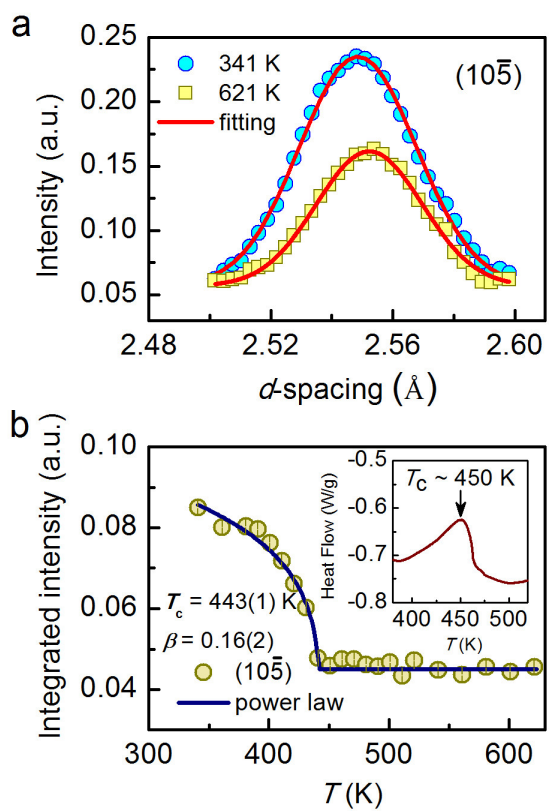

**Figure S1. a**, The Gaussian function fitting of $(10\bar{5})$ Bragg peak of neuron powder diffractions obtained at NOMAD (bank 3) at 341 and 621 K. **b**, The power-law fitting of the temperature dependence of integrated intensity of $(10\bar{5})$ Bragg peak obtained in the Gaussian fitting. The fitting gives rise to that the critical exponent $\beta$ is 0.16(2) and the transition temperature $T_c$ is 443(1) K. This indicates that the critical behavior can be described by two dimensional Ising model. The inset shows the heat flow as a function of temperature, recorded in thermal measurement by using a differential scanning calorimeter. The thermally determined transition temperature is about 450 K, slightly higher than that determined in the power-law fitting.

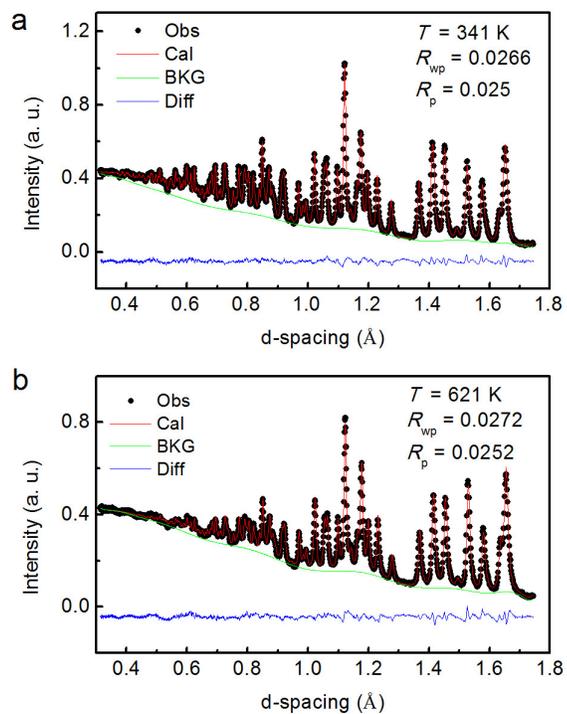

**Figure S2. a**, **b,** Rietveld refinements of neuron powder diffraction patterns obtained from NOMAD (bank 4) at 341 K with $R3m$ model and at 621 K with $R\bar{3}m$ model, respectively. The detailed crystal structure information is summarized in **Table S1** and **S2**.

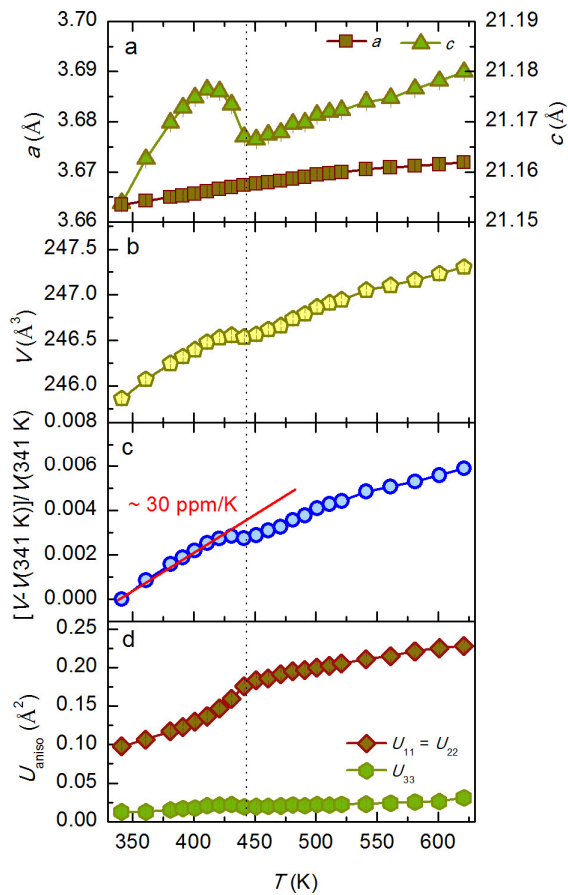

**Figure S3. a, b, c, d,** The lattice dimensions, volumetric thermal expansion, and anisotropic Debye-Waller factors of Ag determined in Rietveld refinements of neutron powder diffraction patterns. Below the transition, the volumetric thermal expansion coefficient is about 30 ppm/K. $T_c$ is labeled by a dot line. Debye-Wall factors of Ag are surprisingly large and significantly anisotropic. $U_{11}$ is almost ten times larger than $U_{33}$, which implies a strong thermal motion confined in the *ab* plane.

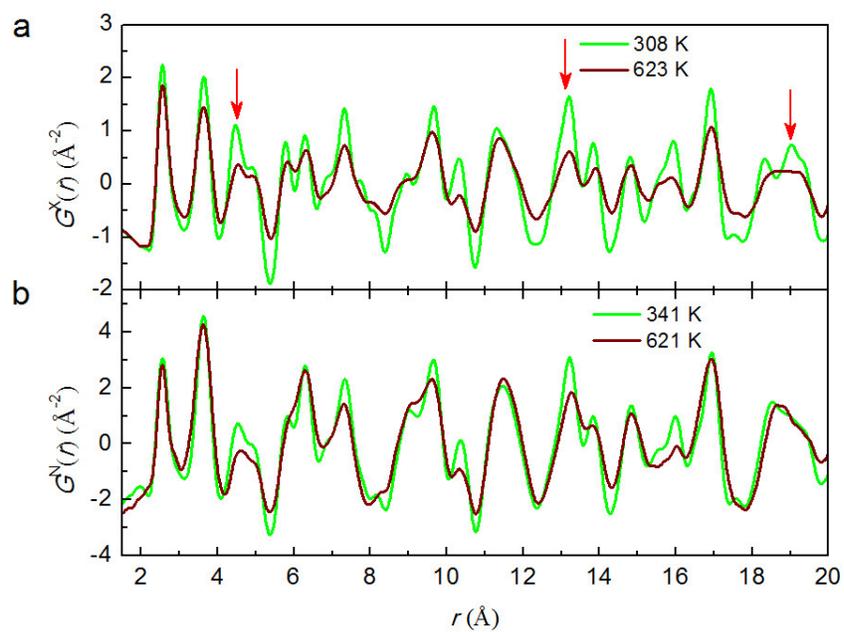

**Figure S4. a, b,** Comparison of experimental $G(r)$ obtained from X-ray scattering and neutron scattering. The arrows mark the Ag-correlation-rich peaks that are much stronger in X-ray case.

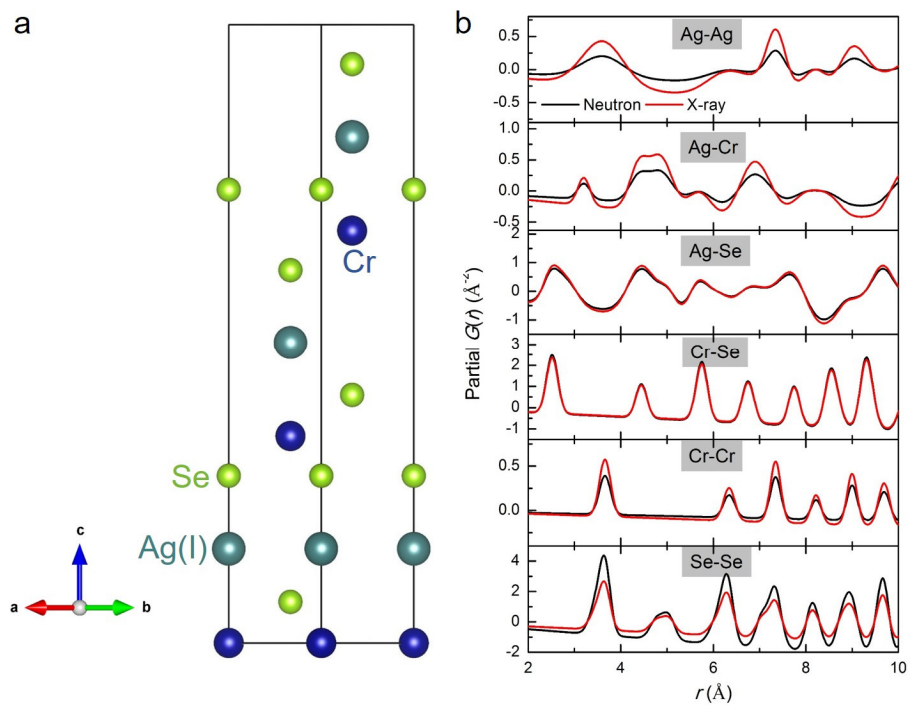

**Figure S5. a**, **b**, The calculated partial $G(r)$ for X-ray scattering and neutron scattering based on the crystal model at 341 K for pairs: Ag-Ag, Ag-Cr, Ag-Se, Cr-Se, Cr-Cr, and Se-Se. The detailed information on this model can be found in **Table S1**. Note that the model used here is disorder free. However, in reality, the occupational disorder of Ag leads to the split of uniform Ag-Ag bond length equal to lattice constant $a$ into three sets of bond lengths (see **Fig. 1b**). The shortest one is equal to the distance between I and II sites. It is about 2.2 Å, but the intensity would be 1/3 of the first Ag-Ag peak shown in **b**. The full width at half maximum of the peak should be constant, because it is related to the Debye-Waller factors. Thus, this short correlation is too weak to see within our experimental resolution.

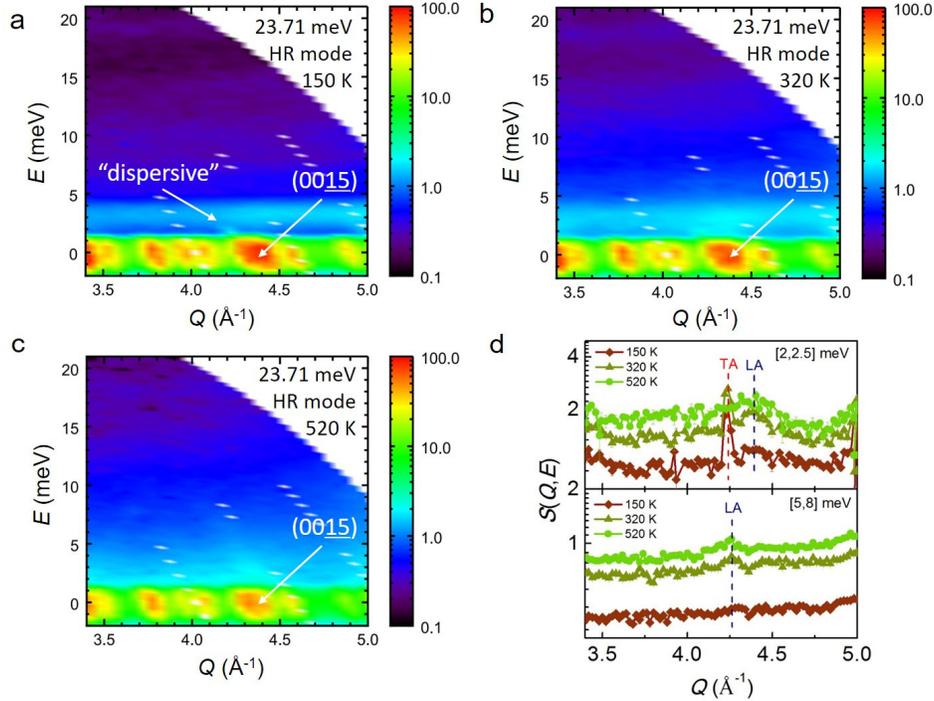

**Figure S6. a, b, c,** $S(Q,E)$ contour maps in the vicinity of the Bragg peak (00$\underline{15}$) obtained with $E_i$ = 23.71 meV in the HR mode at 150, 320, and 520 K. The positions of the Bragg peak (00$\underline{15}$) are labelled. In this region, there is fewer component of diffuse scattering arising from Ag occupational disorder so that it is beneficial to examine the pure information of phonons. At 150 K, dispersive-like intensity is found as the arrow shows, where the damping effect is relatively weak. **d**, The constant energy cuts at [2,2.5] meV (upper panel) and [5,8] meV (lower panel). These cuts give the information of transverse acoustic (TA) phonons and longitudinal acoustic (LA) phonon near "Brillouin zone centers". At 520 K, the peak of TA phonons disappears, but that of LA phonon remains. In concert with the disappearance of peak of TA phonons near Brillouin zone boundaries shown in **Fig. 3** and **4**, it is unequivocally conclusive that TA phonons collapse above $T_c$. Moreover, these dispersive characteristic confirm their nature of acoustic phonons.

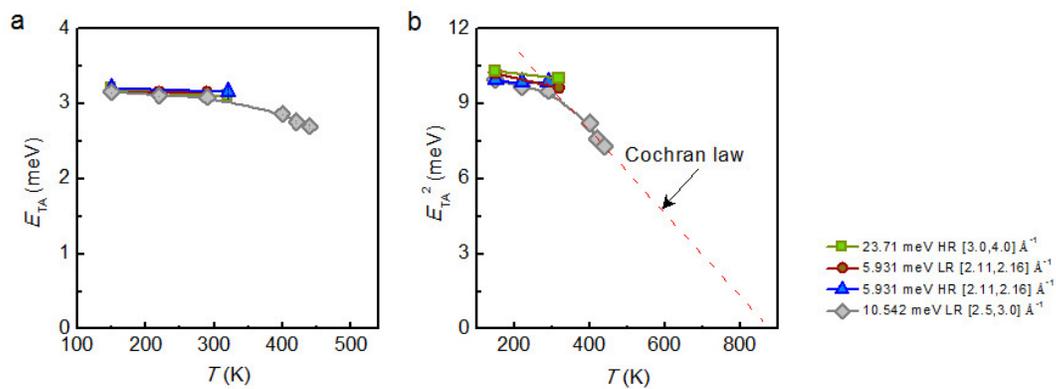

**Figure S7. a**, The energy of TA phonons (at the zone boundaries) obtained at different measurement conditions and they are well consistent with each other. **b**, The square of the energy of TA phonons as a function of temperature. The straight dash line points out the "critical temperature" deduced from Cochran law for a soft-mode transition[S27], which is much higher than the real transition temperature around 450 K. Thus, the observed breakdown of the TA phonons is not the case of the soft-mode transition.

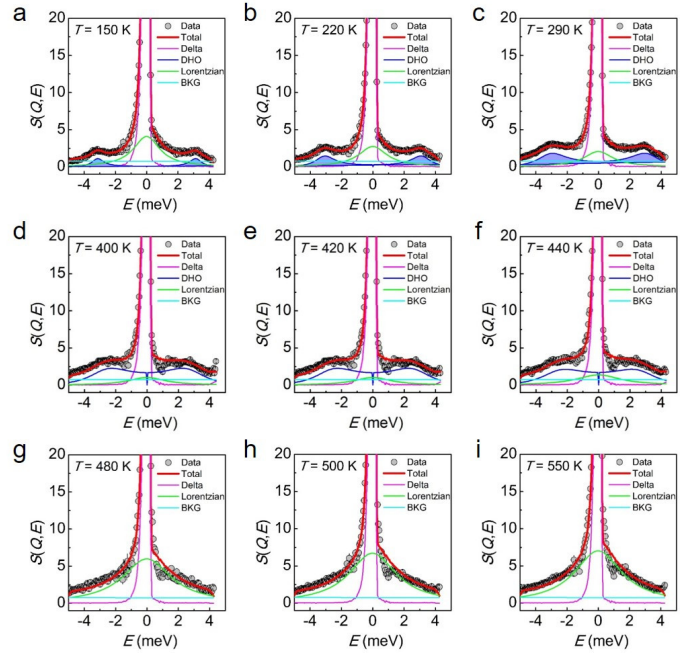

**Figure S8. a – i**, The inelastic neutron scattering spectra ($E_i$ = 5.931 meV in the LR mode) and fitting to a combination of a constant background (BKG), a damped harmonic oscillator (DHO) function, a Lorentzian function and a delta function convoluted by instrumental resolution function below 440 K. At 400, 420, and 440 K, the phonon peaks are too broadened to allow a fitting with free parameters. Thus, the peak position of DHO is fixed at the value determined in 10.542 meV data. The widths of DHO and Lorentzian functions are fixed at appropriate values as well. The small dip around 1 meV of data above room temperature is because of the background subtraction.

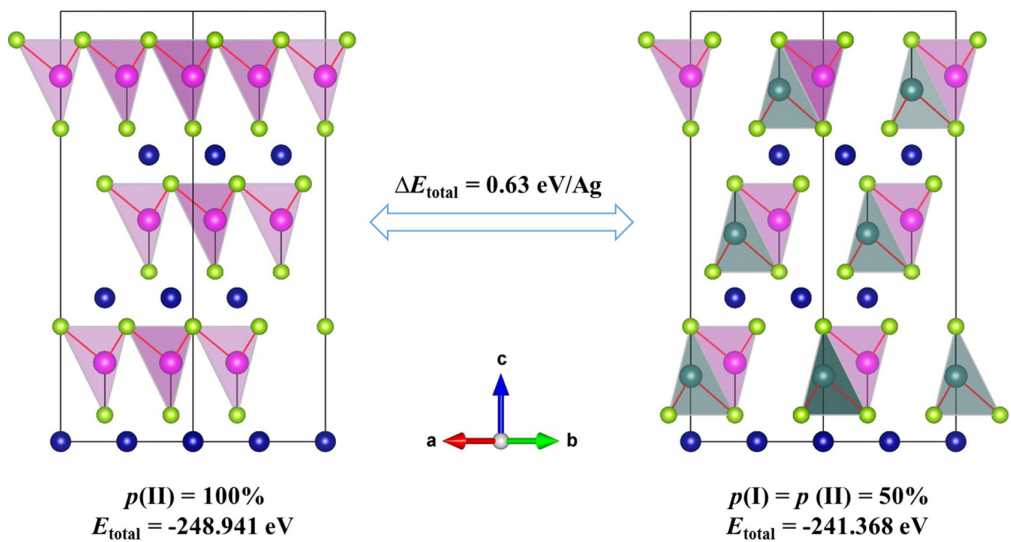

**Figure S9.** The first-principle estimation of energy increase due to occupational disorder of Ag atoms. The left panel is for fully ordered structure with all atoms occupying at II sites. The right panel shows an artificial structure in which half atoms are at I and another half atoms at II. The total energy difference per Ag ion is 0.63 eV.

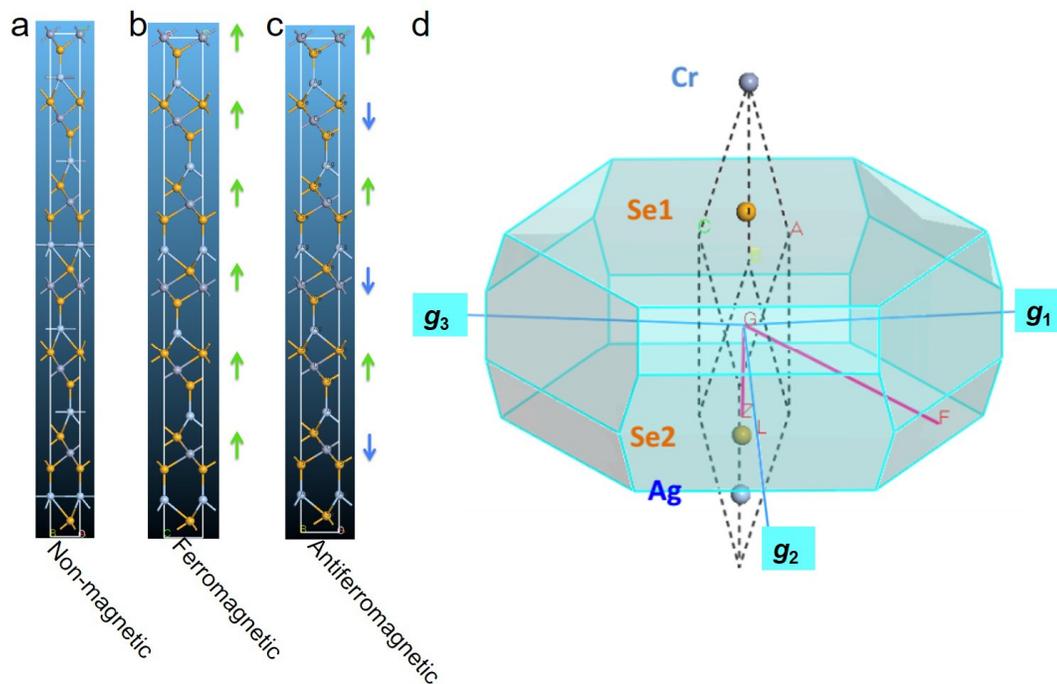

**Figure S10. a**, **b**, **c**, Schematic models of different magnetic configurations for DFT calculations. The lattice dimensions and total energies are summarized in **Table S4**. The lattice dynamics calculation was based on the ferromagnetic configuration. **d**, Definition of high symmetry paths in irreducible Brillouin zone demonstrated by light grey polyhedron. Black dash line and cyan solid line represent the real-space and reciprocal-space lattice, respectively. Blue solid line shows the reciprocal lattice vectors and red solid lines indicate the high symmetry paths.

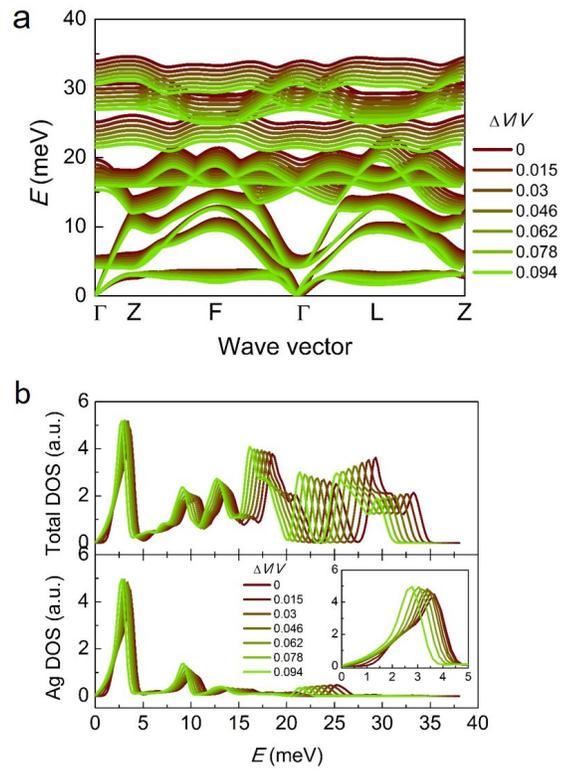

**Figure S11. a**, The phonon dispersions in full frequency range obtained in DFPT QHA calculations. **b**, Total phonon density of state (PDOS) and partial PDOS of Ag. Inset is the zoom-in of the TA phonons.

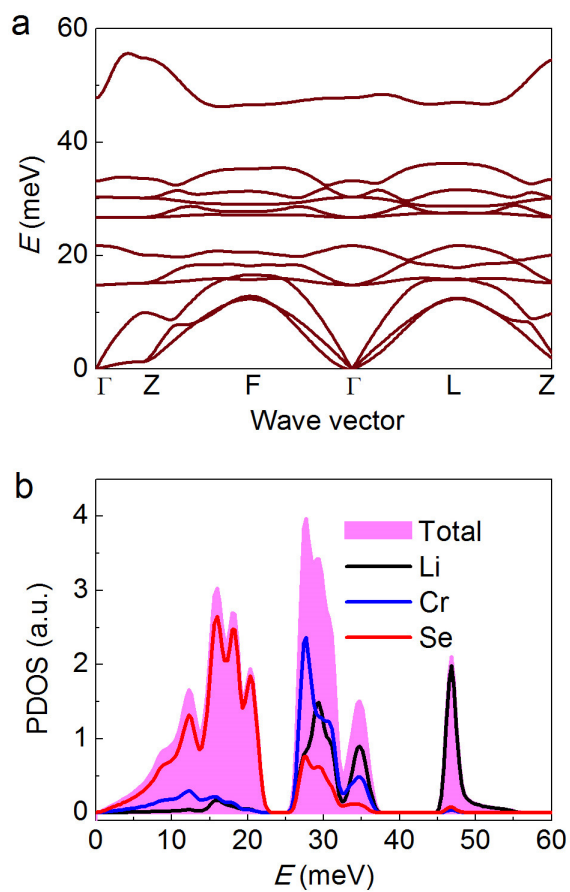

**Figure S12**. **a**, The phonon dispersion of LiCrSe$_2$ in same structure with AgCrSe$_2$ in *R*3*m* symmetry. **b**, The total and partial PDOS. The optimized structure and total energy are listed in **Table S4**.

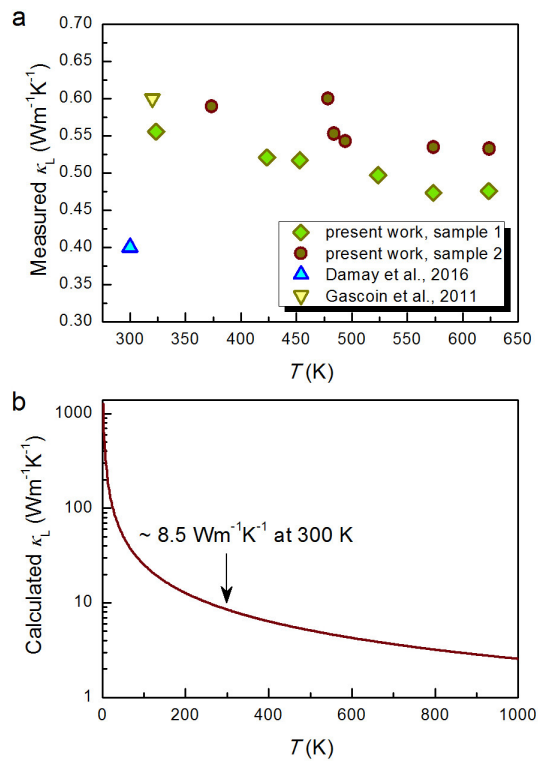

**Figure S13. a**, Experimentally measured thermal conductivity as a function of temperature for two different samples of $AgCrSe_2$. The thermal conductivity data from references (*4*) and (*5*) are also listed for comparison. **b**, The calculated temperature dependent thermal conductivity by DFPT QHA method. Such huge difference between experimental and calculated values are because the disorder effect is not taken into account in the DFPT QHA calculation.

3. Supplemental tables

Table S1. The crystal structure data at 341 K. The space group is *R3m*. $a$ = 3.66339(7) Å, $c$ = 21.1537(6) Å. $R_{wp}$ = 0.0266, $R_p$ = 0.025.

| Atom | (0, 0, z) | $U_{11} = U_{22} = 2U_{12}$ (Å²) | $U_{33}$ (Å²) | $U_{13} = U_{23}$ (Å²) |
|---|---|---|---|---|
| Ag | 0.15184(10) | 0.0977(13) | 0.0123(14) | 0 |
| Cr | 0.00018(34) | 0.0074(5) | 0.0052(9) | 0 |
| Se1 | 0.26957(16) | 0.00673(26) | 0.0115(4) | 0 |
| Se2 | 0.73310(13) | 0.00673(26) | 0.0115(4) | 0 |

Table S2. The crystal structure data at 621 K. The space group is $R\bar{3}m$. $a$ = 3.67187(10) Å, $c$ = 21.1799(8) Å. $R_{wp}$ = 0.0272, $R_p$ = 0.0252.

| Atom | (0, 0, z) | $U_{11} = U_{22} = 2U_{12}$ (Å²) | $U_{33}$ (Å²) | $U_{13} = U_{23}$ (Å²) |
|---|---|---|---|---|
| Ag | 0.15149(21) | 0.228(5) | 0.0309(32) | 0 |
| Cr | 0 | 0.0131(9) | 0.0171(14) | 0 |
| Se | 0.26808(5) | 0.0147(6) | 0.0220(7) | 0 |

Table S3. The comparison of optimized structures with 100% I and 100% II occupations in *R3m* symmetry.

|  | I | II |
|---|---|---|
| **Total energy (eV)** | -62.211734 | -62.211518 |
| **$a = b$ (Å)** | 3.753568 | 3.753363 |
| **$c$ (Å)** | 21.102433 | 21.104723 |
| **$\alpha = \beta$ (°)** | 90 | |
| **$\gamma$ (°)** | 120 | |
| **Ag coordinate** | (0, 0, 0.1502689) | (2/3, 1/3, 0.18426765) |
| **Cr coordinate** | (0, 0, 0.99988091) | (0, 0, 0.00138861) |
| **Se1 coordinate** | (0, 0, 0.27152302) | (0, 0, 0.26830460) |
| **Se2 coordinate** | (0, 0, 0.73302718) | (0, 0, 0.72976914) |

**Table S4.** The optimized lattice constants and total energies in DFT calculations on a twice supercell along the *c* axis. There are totally 24 atoms in this supercell.

|  | Non-magnetic | Ferromagnetic | Antiferromagnetic | LiCrSe$_2$ |
|---|---|---|---|---|
| **Total energy (eV)** | -118.462 | -124.472 | -124.478 | -129.695 |
| ***a = b* (Å)** | 3.274 | 3.753 | 3.749 | 3.7219 |
| ***c* (Å)** | 48.902 | 42.201 | 42.239 | 40.6082 |
| ***α = β* (°)** | 90 | | | |
| ***γ* (°)** | 120 | | | |

**Table S5.** Group velocities of transverse acoustic (TA) and longitudinal acoustic (LA) phonons determined from the calculated dispersion data shown in Fig. S11.

| Direction | $v_{TA1}$ (m s$^{-1}$) | $v_{TA2}$ (m s$^{-1}$) | $v_{LA}$ (m s$^{-1}$) |
|---|---|---|---|
| Γ-Z | 940 | 940 | 3092 |
| Γ-F | 998 | 1244 | 2373 |
| Γ-L | 1004 | 1163 | 2466 |
| **Average** | 981 | 1116 | 2644 |

## 4. Supplemental references